\newcommand{\beq}{\begin{equation}}
\newcommand{\eeq}{\end{equation}}
\newcommand{\bea}{\begin{eqnarray}}
\newcommand{\eea}{\end{eqnarray}}

\newcommand{\gsim}{\lower.7ex\hbox{$\;\stackrel{\textstyle>}{\sim}\;$}}
\newcommand{\lsim}{\lower.7ex\hbox{$\;\stackrel{\textstyle<}{\sim}\;$}}

\newcommand{\mrm}{\mathrm}

\documentclass[aps,prev,twocolumn,preprintnumbers,floatfix,nofootinbib]{revtex4-1}
\usepackage{graphicx}
\usepackage{epstopdf}
\usepackage{mathrsfs}
\usepackage{amssymb}
\usepackage{verbatim}
\usepackage{color}
\usepackage{multirow}

\def\stacksymbols #1#2#3#4{\def\theguybelow{#2}
    \def\vp{\lower#3pt}
    \def\sp{\baselineskip0pt\lineskip#4pt}
    \mathrel{\mathpalette\intermediary#1}}

\def\intermediary#1#2{\vp\vbox{\sp
     \everycr={}\tabskip0pt
     \halign{$\mathsurround0pt#1\hfil##\hfil$\crcr#2\crcr
              \theguybelow\crcr}}}

\def\be{\begin{equation}}
\def\ee{\end{equation}}
\def\bea{\begin{eqnarray}}
\def\eea{\end{eqnarray}}

\def\sp{\;\;\;,\;\;\;}

\def\mrm{\mathrm}

\def\lsim{\raise0.3ex\hbox{$\;<$\kern-0.75em\raise-1.1ex\hbox{$\sim\;$}}}
\def\gsim{\raise0.3ex\hbox{$\;>$\kern-0.75em\raise-1.1ex\hbox{$\sim\;$}}}

\def\inbar{\,\vrule height1.5ex width.4pt depth0pt}

\def\IC{\relax\hbox{$\inbar\kern-.3em{\rm C}$}}
\def\IQ{\relax\hbox{$\inbar\kern-.3em{\rm Q}$}}
\def\IR{\relax{\rm I\kern-.18em R}}
 \font\cmss=cmss10 \font\cmsss=cmss10 at 7pt
\def\IZ{\relax\ifmmode\mathchoice
 {\hbox{\cmss Z\kern-.4em Z}}{\hbox{\cmss Z\kern-.4em Z}}
 {\lower.9pt\hbox{\cmsss Z\kern-.4em Z}}
 {\lower1.2pt\hbox{\cmsss Z\kern-.4em Z}}\else{\cmss Z\kern-.4em Z}\fi}

\def\comment#1{}
\def\to{\rightarrow}

\def\u1x{U(1)_X}
\newcommand{\nc}{\newcommand}
\nc{\LL}{L}
\nc{\vv}{\tilde{v}}
\nc{\ccdot}{\!\cdot\!}
\nc{\gsm}{G_{SM}}
\nc{\vfive}{\mathbf{5}\oplus\mathbf{\overline{5}}}
\nc{\vten}{\mathbf{10}\oplus\mathbf{\overline{10}}}
\nc{\zhol}{Z^{\rm hol}}
\nc{\xfb}{\,{\rm fb}}

\begin{document}

%
%

\preprint{LPT--Orsay 14-18}
\preprint{CPHT-RR016.0414}

\vspace*{1mm}

\title{Generating X-ray lines from annihilating dark matter }

\author{Emilian Dudas$^{a}$}
\email{emilian.dudas@cpht.polytechnique.fr}
\author{Lucien Heurtier$^{a}$}
\email{lucien.heurtier@cpht.polytechnique.fr}
\author{Yann Mambrini$^{b}$}
\email{yann.mambrini@th.u-psud.fr}

\vspace{0.1cm}
\affiliation{
${}^a$ CPhT, Ecole Polytechnique, 91128 Palaiseau Cedex, France }
\affiliation{
${}^b$ 
Laboratoire de Physique Th\'eorique, 
Universit\'e Paris-Sud, F-91405 Orsay, France 
}

\begin{abstract} 

We propose different scenarios where a keV dark matter annihilates to produce a monochromatic signal.
 The process is generated through the exchange of a light scalar of mass of order 300 keV - 50 MeV coupling
to photon through loops or higher dimensional operators. For natural values of the couplings and scales, the model can generate
 a gamma-ray line which can fit with the recently identified 3.5 keV X-ray line. 
  
\end{abstract}

\maketitle


\maketitle


\setcounter{equation}{0}



\section{Introduction}

\noindent
Physics community is still in  the quest of the dominant matter component in the Universe. 
Even if we know quite well the cosmological abundance of this dark matter \cite{WMAP, PLANCK}, little is known
about its mass and couplings. In view of recent results from both direct and indirect detection experiments,
the Weakly Interacting Massive Particle (WIMP) paradigm, corresponding to a $\sim 100$ GeV particle interacting weakly 
with the visible sector 
begins to be severely constrained by XENON \cite{XENON}, LUX \cite{Akerib:2013tjd} or FERMI satellite \cite{Ackermann:2013yva}. 
 On the other hand, several other scenarios
offer much lighter \cite{Kusenko:2009up} or heavier \cite{Kolb:1998ki} candidates with feeble \cite{freezein1,freezein2} or 
very feeble \cite{Mambrini:2013iaa} couplings. 
Their thermal histories are relatively different (but not less motivated) from the standard freeze out one.
The cases of FIMP (for  Freeze In Massive Particle or Feebly Interacting Massive Particle) or WISP (for
Weakly Interacting Slim Particle) are typical cases where the coupling is too weak to reach the thermal
equilibrium with the standard model bath \cite{freezein1}. The dark matter candidate can be so weakly coupled that
it decoupled from the bath at the reheating epoch, like the gravitino or candidates motivated by 
SO(10) schemes. Other scenario proposed an even more weakly interacting particle, so weakly interacting that
the dark matter is stable at the scale of the age of the universe: the decaying dark matter (see \cite{Dugger:2010ys} for a review
on the subject). 

\noindent
At present, clues for the presence of an interacting dark matter are rare. However, recently a 3.55 keV X-ray
line has been reported in the stacks analysis of 73 galaxy clusters from the XMM-Newton telescope \cite{Bulbul:2014sua} 
with a significance larger than 3$\sigma$.
A similar analysis finds evidence at the 4.4 $\sigma$ level for a 3.52 keV line from their analysis of the X-ray spectrum
 of the Andromeda galaxy (M31) and the Perseus Cluster \cite{Boyarsky:2014jta}. In both analysis, the unidentified line was interpreted 
 as a possible signal of sterile neutrino dark matter $\nu_s$ \cite{sterile} decaying through a loop $\nu_s \rightarrow \gamma \nu$.
 While more conventional explanations in term of atomic physics effects are currently lacking, 
  several works have been released in the following weeks, all focussing on a decaying dark matter candidate.
  Extensions with a sterile neutrino as dark matter candidate \cite{sterilebis}, axions or ALPs \cite{axion}, axinos \cite{axino}, pseudo-Nambu-Goldstone bosons \cite{Nakayama:2014cza} or
 supersymmetric models (gravitino \cite{gravitino}, sgoldstino \cite{demidov}
or low scale supersymmetry breaking \cite{Ko:2014xda}) were proposed, all relying on processes ensuring a fine tuned lifetime $\tau \simeq 10^{28}$ seconds to fit the
  observed line. Other more exotic candidates like decaying moduli \cite{moduli}, millicharged dark matter \cite{milli}, dark atoms\footnote{For an introduction to the subject, see e.g. \cite{darkatomsreview}} \cite{darkatoms}, magnetic dark matter \cite{magnetic}, majoron decay \cite{majoron} or
  multicomponent \cite{multi} have been proposed.
 Another original model 
 was studied in  \cite{Finkbeiner:2014sja, eXciting} with an eXciting dark matter, where the photons come from the transition from
 an excited state down to the ground state for the dark matter particle, which in this case can be significantly heavier than 3.5 keV. Moreover, it is well known that for a warm dark matter candidate of mass $\sim$keV, free streaming produces a cutoff in the linear fluctuation power spectrum at a scale
corresponding to dwarf galaxies and can fit observations for $m_s \gtrsim 1.5$ keV \cite{Lovell:2013ola}.
  
  \noindent
  All of these scenarios seem to exclude (maybe too early) an annihilating dark matter scenario \cite{Frandsen:2014lfa}.
Indeed, the cross section needed to fit the excess
 measured in \cite{Bulbul:2014sua, Boyarsky:2014jta} is $\sigma v \simeq 10^{-33} ~\mrm{cm^3 s^{-1}}$ which corresponds to an effective scale $\Lambda$
 of $ \Lambda \simeq 10$ GeV for a 3.5 keV dark matter, in the classical effective operators approach. This scale is too low to be due to heavy 
 particles charged under the electromagnetic field.
 
 \noindent
  In this work however, we show that such cross section arises naturally when we go beyond the naive
  effective operators approach and consider microscopic constructions with the presence of a light
 hidden mediator coupling to the standard model through effective operators. Such approach had been explored in the case of a vector boson or a heavy fermionic mediator \cite{boehm1} but did not consider the possibility of a scalar one. Furthermore these studies had worked out a lower bound on the dark matter mass of order $\mathcal O \left(\mrm{MeV}\right)$ in order to fit both ray fluxes and relic density measurements \cite{boehm2}. We will show that this bound can actually be overcome if one considers the dark sector to be living in a thermal bath of temperature differing from the one of the visible sector.

\noindent
The paper is organized as follows. After a summary 
 of the effective operators approach in the next section, we describe our model in  Section III and
 extract the parameter space allowed to fit the observed X-ray line.  We then compute
 the relic abundance predicted in this restricted parameter space by adding an additional coupling
of the mediator to a light sterile right-handed neutrino in Section IV.  We show that the model can  fit WMAP/PLANCK data
 taking into account that the bath of the hidden sector is at a different temperature compared to the standard model one. 
 We then conclude in section V with an explicit example of an UV  model, before giving the relevant technical formulae in the appendix in the case of a fermionic dark matter.
 


\section{The failure of a naive effective  operators approach}
\label{Sec:effective}

\subsection{A priori}
\noindent
Since Fermi's time the classical way to explain a signal or to work out predictions at a scale beyond the standard model
one\footnote{The  "standard model scale" varying with time: $\simeq$ MeV in 1897, $\simeq$ GeV in 1932, $\simeq$ 100 GeV
in 1983, TeV scale nowadays.} (which is the one that one can reach experimentally at a given time) is to use the effective operators approach.
Indeed, it seems natural to imagine that the physics beyond a measurable scale of energy is represented by heavy
particles not yet produced in accelerators, but coupling to the observable sector. This is frequently used in LHC studies
\cite{Goodman:2010ku} or dark matter searches \cite{Mambrini:2011pw}. 
However, it is obvious that this effective operators approach has its limitations. First of all, the coupling to the visible
sector generated by loops of heavy states can be highly dependent on the hidden microscopic physics 
(gauge or Yukawa--like couplings) but, more dramatically, the presence of light states modifies considerably the 
predictions, as was shown in \cite{Papucci:2014iwa}. Supersymmetric or grand unified models do not escape this rule:  light stau 
or $Z'$ for instance generate new processes observable at LHC and not predicted by a naive effective operators approach.
This is exactly what is happening in the case of a cosmological monochromatic signal as one discusses below.

\subsection{The monochromatic signal}
\noindent
If the signal analysed in \cite{Bulbul:2014sua,Boyarsky:2014jta} is generated by dark matter annihilation
to two photons $s s \rightarrow \gamma \gamma$ (with a  dark matter mass of 3.5 keV) then one should
 fit the annihilation cross section
$\langle \sigma v \rangle_{\gamma \gamma}$ with the flux measured in the vicinity of the sun.
A naive estimate of the total luminosity of Perseus can be computed using

\be
L = \int_{0}^{R_{Pe}} 4 \pi r^2 n^2_{DM} \langle \sigma v \rangle_{\gamma \gamma}=
\int_{0}^{R_{Pe}} 4 \pi r^2 \left( \frac{\rho(r)}{m_s} \right)^2 \langle \sigma v \rangle_{\gamma \gamma}
\label{Eq:luminosity}
\ee

\noindent
with $R_{pe}$ is the Perseus radius. As a first approximation, one can consider like in \cite{Krall:2014dba} a mean density
of dark matter in the cluster. The Perseus observation involved a mass\footnote{See table 2 of \cite{Bulbul:2014sua}} 
of $M_{Pe}= 1.49 \times 10^{14} M_{\odot}$ in a region of $R_{Pe}=0.25$ Mpc at a distance of $D_{Pe}=78$ Mpc from  the solar system.
One can then estimate

\begin{eqnarray}
n_{DM} \simeq \frac{1.49 \times 10^{14}M_{\odot} }{m_s} \frac{3}{4 \pi R_{Pe}^3} &=& 2.0\times 10^{-37}\mrm{GeV^3} \nonumber \\
&=& 2.6 \times 10^4 \mrm{cm^{-3}} \ .
\label{Eq:ndm}
\end{eqnarray}

\noindent
Combining Eq.(\ref{Eq:luminosity})  and (\ref{Eq:ndm}), one can compute the luminosity in the Perseus cluster,

\be
L \simeq 1.2 \times 10^{55} \left( \frac{3.5 ~\mrm{keV}}{m_s} \right)^2
 \left( \frac{\langle \sigma v \rangle_{\gamma \gamma}}{10^{-26} \mrm{cm^3 s^{-1}}} \right) ~~\mrm{ph/s}
\ . \ee

\noindent
One can then deduce the flux $\phi_{\gamma \gamma} = L/(4 \pi D_{Pe}^2)$ that one should observe on earth

\be
\phi_{\gamma \gamma} = 1.7 \times 10^{-5} \left( \frac{3.5 ~ \mrm{keV}}{m_s} \right)^2 
\left( \frac{\langle \sigma v \rangle_{\gamma \gamma}}{10^{-32} \mrm{cm^3 s^{-1}}} \right)~ \mrm{cm^{-2} s^{-1}}
\ee

\noindent
The cumulative flux of $\phi_{\gamma \gamma} \sim 4 \times 10^{-6} \mrm{cm^{-2} s^{-1}}$ from \cite{Bulbul:2014sua} is hard to interpret
in the dark matter framework as it arises from a combination of clusters at a variety of distances. However, according to
the authors of \cite{Boyarsky:2014jta,Finkbeiner:2014sja}, one can identify a monochromatic signal arising from M31 or Perseus cluster with a flux {\color{blue}} $\phi_{\gamma \gamma}=5.2_{-2.13}^{+3.70} \times 10^{-5}$ photons $\mrm{cm^{-2}}$ per seconds at
3.56 keV with the cluster core. 
One can make a more refined analysis, considering for instance a more complex halo structure like Einasto 
or NFW profile as in \cite{Finkbeiner:2014sja}
but the result will not change dramatically\footnote{We thank M. Yu. Khlopov for having drawn our attention to ref. \cite{Belotsky:2014doa} for a detailed analysis on profiles concerning gamma ray production from so called dark matter clumps}. One could also look at the Centaurus observation like in 
\cite{Krall:2014dba} with $M_{Ce} = 6.3 \times 10^{13} M_{\odot}$ and a radius of $R_{Ce} = 0.17$ Mpc.

\noindent
Finally, taking into account also other observations like M31, we will impose in our analysis a conservative required  annihilation cross section estimated as

\be
\langle \sigma v \rangle_{\gamma \gamma} \simeq (2 \times 10^{-33} - 4 \times 10^{-32}) \ \mrm{cm^3 s^{-1}} \ . 
\ee

\noindent
However, for such a light dark matter particle annihilating into photons, it is important to check the consequences
of injecting secondary particles on the recombination, leaving an imprint on Cosmic Microwave Background
(CMB) anisotropies. The authors of \cite{Lin:2011gj} show that the corresponding condition is given by

\be
\langle \sigma v \rangle_{\gamma \gamma}^{CMB} \lesssim 2.42 \times 10^{-27 } \left( \frac{m_s}{1~\mrm{GeV}} \right) ~\mrm{cm^3 s^{-1}} \ , 
\ee

\noindent
which for a 3.5 keV dark matter is $\langle \sigma v \rangle_{\gamma \gamma} <  8.5 \times 10^{-33} \mrm{cm^3 s^{-1}} $.  $In$ $fine$
one will  then restrict ourself to the parameter space allowing a monochromatic signal and respecting the CMB
constraints:

\be
2 \times 10^{-33} \mrm{cm^3 s^{-1}} < \langle \sigma v \rangle_{\gamma \gamma} < 8.5 \times 10^{-33} \mrm{cm^3 s^{-1}} \ . 
\label{Eq:constraints}
\ee
\subsection{A posteriori}

\begin{figure}
\begin{center}
 \includegraphics[width=0.5\linewidth]{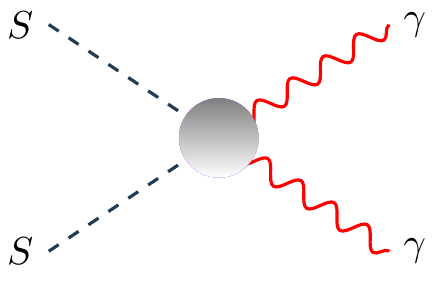}
 \caption{{\footnotesize Effective diagram for dark matter annihilation}}
\label{Fig:feynmanscalar}
\end{center}
\end{figure}

\noindent
In the case of a scalar particle annihilating into two photons, the CP--even effective lagrangian
can be written\footnote{To simplify the analysis, we will consider a scalar
dark matter candidate with CP-even couplings thorough the paper. The other cases (fermionic dark matter, 
CP-odd or pseudo--scalar couplings..) change our conclusion by factors of order of unity and are 
treated in appendix.}  

\be
{\cal L}_{eff} = \frac{S^2}{\Lambda^2} F_{\mu \nu} F^{\mu \nu} \ , 
\ee

\noindent
with $F_{\mu \nu} = \partial_\mu A_\nu - \partial_\nu A_\mu$ 
being the electromagnetic field strength.
 The scale $\Lambda$ is related to the mass of the particles running in the loops (see Fig.(\ref{Fig:feynmanscalar})) which , being charged
under $U(1)_{em}$,  should be heavier or at least of the order of  TeV. A list of generic couplings of this type can be found in
 \cite{Chu:2012qy}. We will write in the appendix the results we  obtained in other cases.

\noindent
It is important to notice that such a light dark matter can contribute to the effective number of neutrinos $N_{eff}$. 
However, it has been shown recently that a dark matter annihilating into photons in sub-MeV masses is possible only in the
case of a scalar dark  matter \cite{Boehm:2013jpa}.
Such processes have already been computed in \cite{Rajaraman:2012fu} and one obtains

\be
\langle \sigma v \rangle_{\gamma \gamma}^{eff} = \frac{2 m_s^2}{\pi \Lambda^4} \ . 
\label{Eq:sigmaeffective}
\ee

\noindent
Applying the constraints (\ref{Eq:constraints}) to the annihilation cross section (\ref{Eq:sigmaeffective})
one obtains

\be
10~\mrm{GeV} < \Lambda < 15 ~\mrm{GeV} \ . 
\label{Eq:resultseffective}
\ee

\noindent
This value is obviously far below any accelerator limit on charged particles. It seems then impossible to UV complete
this operator and achieve a large enough rate. However, as we will see below, the effective operators approach cannot be applied anymore when the UV sector contains light states.

\section{A natural microscopic approach}\label{micro}

\noindent
It is then natural to build a microscopic model and to see how observables are modified. 
But natural in which sense? Natural in the sense that the presence of a keV-MeV dark matter particle
$naturally$ leads to a keV scale dynamics, as the presence of GeV particles in the standard model
naturally leads to GeV scale dynamics in the Higgs sector.  We then can suppose the presence of a 
(pseudo)scalar coupling to the dark matter candidate, and generating the keV dynamics. The simplest
way to generate such dynamics is through a "higgs-like" portal. We will consider by simplicity 
a scalar dark matter; other dark matter spin or couplings do not change our conclusions
and are treated in the appendix.

\subsection{The scalar model}\label{model}

\noindent
We will work in the framework of a scalar portal $\phi$, coupling directly at tree-level to dark matter,
but indirectly to the standard model through loops. This is a typical secluded dark matter type of model \cite{Pospelov:2007mp}.
The lagrangian can then be written for a scalar dark matter

\be
{\cal L}_{eff}\supset -\frac{m_s^2}{2} S^2 -\frac{m_{\phi}^2}{2} \phi^2 - \tilde m  \phi S^2 + \frac{\phi}{\Lambda} F_{\mu \nu} F^{\mu \nu}.
\label{Eq:lagrangian}
\ee  

\noindent
We assume the parameter $\tilde m$ to be a free mass scale parameter. However such coupling can be explicitely generated by symmetry breaking in renormalizable models, as illustrated in section \ref{examples}. In the latter case, $\tilde m$ is expected to be at most of the same order of magnitude than $m_{\phi}$ since it gets its value from the $vev$ of a field $\Phi=v_{\phi}+\phi$ after spontaneous symmetry breaking. Furthermore this is what would be more generally expected if $\tilde m$ is generated by whatever dynamical mechanism involving only $\phi$ and the light field $S$. The mass scale $\Lambda$ is related to the mass of heavy particles integrated in the loop. In a perturbative set up with $N$ charged fermions running in the loop $\Lambda \sim \frac{4\pi}{N h_{\phi} \alpha}M_{\psi}$, where $h_{\phi}$ is the Yukawa coupling of $\phi$ to the charged fermions of mass $M_{\psi}$. Using the constraint  $M_{\psi} \gtrsim 500~\mrm{GeV}$ from collider searches and perturbativity one finds that the minimum 
natural values for $\Lambda$ are $\Lambda \sim 50-500~\mrm{TeV}$, whereas $\Lambda \sim 5~\mrm{TeV}$ can only be obtained in a strongly coupled hidden sector.

\begin{figure}
\begin{center}
 \includegraphics[width=1.05\linewidth]{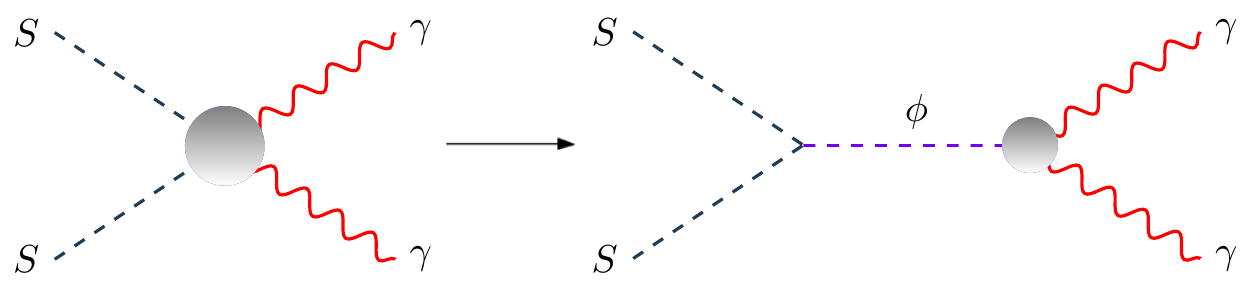}
 \caption{{\footnotesize Microscopic diagram for dark matter annihilation}}
\label{Fig:feynmanmicro}
\end{center}
\end{figure}

\noindent
Such a lagrangian gives for the annihilation cross section (process depicted in Fig.(\ref{Fig:feynmanmicro}) )

\be
\langle \sigma v \rangle_{\gamma \gamma}^{micro} = \frac{4 m_s^2 \tilde m^2}{\pi \Lambda^2 (4 m_s^2 - m_\phi^2)^2}\, \ .
\ee

\subsection{X-ray line}

\noindent
Depending on the hierarchy between the masses of the mediator $\phi$ and the dark matter particle $S$, the condition (\ref{Eq:constraints}) leads to two kinds of constraints :

\bea
\text{{\bf Case A}} &:&  ~m_\phi \gtrsim m_s~\text{{(Heavy Mediator)}} , 
\nonumber
\\
\nonumber\\
m_\phi \simeq (12.3 &-& 17.6)\sqrt{ \frac{m_s}{3.5 ~\mrm{keV}} } \sqrt{\frac{\tilde m}{\Lambda}} ~\mrm{GeV}
\nonumber
\\
\label{Eq:heavy}\\
\text{{\bf Case B}} &:& ~ m_\phi \lesssim m_s~\text{{(Light Mediator)}},
\nonumber
\\
\nonumber\\
\frac{\tilde m}{\Lambda} &\sim& (1.63-3.36)\times  10^{-13}\ . 
\label{Eq:line}
\eea

\noindent
Both cases give at first sight viable results.
One can understand easily why it is so in the microscopic approach compared to the effective operators approach of Eq.(\ref{Eq:sigmaeffective}).
Indeed, as recently emphasized by the authors of \cite{Papucci:2014iwa} for the LHC analysis of mono jet events, the effective  operators approach
ceases to be valid once the ultraviolet (microscopic) theory contains some light mediators, which is exactly our case. This
comes from two powers less in $\Lambda$ in the computation of observables: heavier states become now reasonably heavy compared
to the result Eq.(\ref{Eq:resultseffective}). 

\noindent
We will see however that experimental bounds on light scalar particle interactions 
with the electromagnetic sector are strongly restrictive.

%
\subsection{Experimental Bounds}\label{sec:exp}
\noindent
As we just mentioned above, interactions of a light scalar, or axion-like particle (ALP) with the visible sector
is very much constrained by collider data (LEP) and astrophysics. 
Indeed bounds on pseudoscalar particles interacting with photons (see \cite{Kleban:2005rj}) have been studied, using
 LEP data from ALEPH, OPAL, L3 and DELPHI, and shown that the coupling of the pseudoscalar with photons cannot 
 exceed a value of $2.6 \times 10^{-4} \mrm{GeV}^{-1}$ for a mediator of mass $m_{\phi}~ \lesssim ~50 ~\mrm{MeV}$, which means, 
 in terms of our mass scale
\be
\Lambda ~\gtrsim ~3~ \mrm{TeV}~~~~~[m_{\phi}~ \lesssim ~50 ~\mrm{MeV}]\,.
\ee
Furthermore, one of the most restrictive constraints on ALPs comes from the non-observation of anomalous energy loss of
 horizontal branch (HB) stars via a too important ALP production \cite{Raffelt:1996wa}. Indeed those contraints impose
\be
\Lambda ~\gtrsim ~10^{10}~ \mrm{GeV}~~~~~[m_{\phi}~ \lesssim ~30 ~\mrm{keV}]\,, 
\ee
for a mediator mass up to $ m_{\phi}\lesssim 30 ~\mrm{keV}$. At higher masses arise constraints coming from the CMB and 
BBN studies, setting lower limits on the coupling with photons. A nice review on the subject can be found in \cite{review1,review2}. 
Various astrophysical constraints on ALP mass and coupling to photons are summarized in, e.g. \cite{Hewett:2012ns}. 

\noindent
These constraints on our model essentially put lower bounds on $\Lambda$. Indeed, for a light mediator (Case B) HB experiments impose that the mass scale $\Lambda$ takes very high values ($\gtrsim 10^{10}~\mrm{GeV}$). In this case, as indicated by Eq.(\ref{Eq:line}), one would need the tri-linear coupling to be of order $\tilde m \gtrsim 10^{-3} ~\mrm{GeV}$. However, in this case, since $m_{\phi}$ is assumed to be smaller than the keV scale, one would conclude that $\tilde m/m_{\phi} \gtrsim 10^3$ which is, as mentioned in the previous section, quite unnatural. We will then concentrate our study on Case A, where the mediator  $\phi$ is assumed to be heavier than the dark matter field $S$.

\begin{figure}[htbp]
 \begin{center}
 \includegraphics[width=0.95\linewidth]{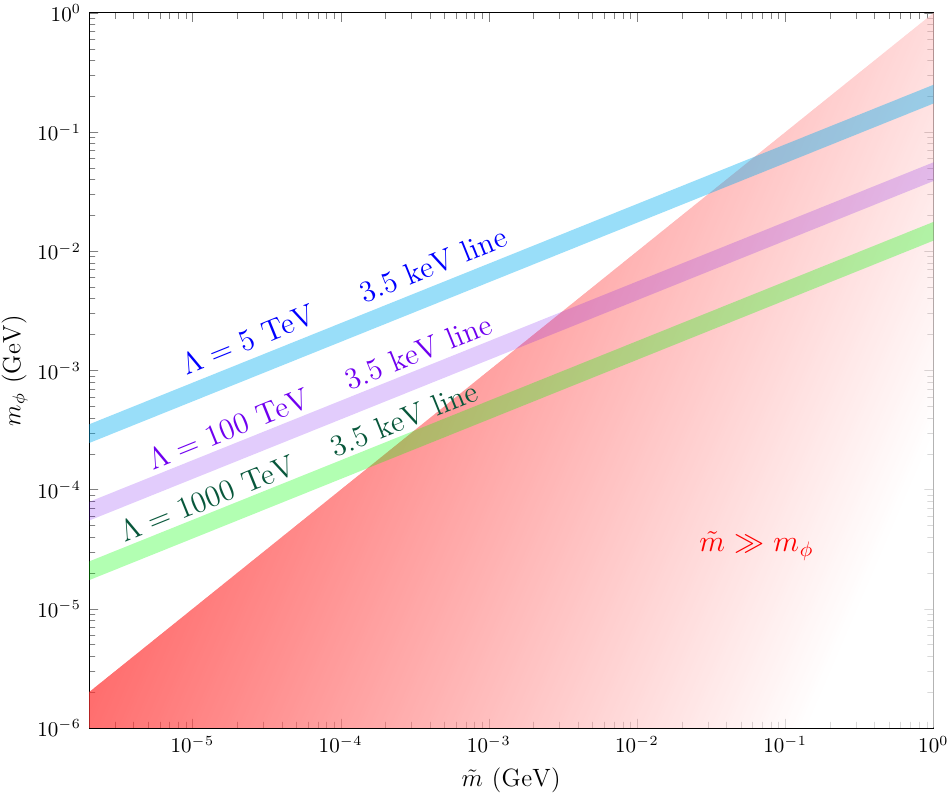}
 \caption{{\footnotesize \label{Fig:CaseA} ($m_{\phi}$,$\tilde m$) parameter space allowed by the $\gamma$ flux measurements in the case of a heavy mediator (Case A), for different values of $\Lambda$. The red shaded region indicates where $\tilde m$ is higher than $m_{\phi}$.}}
\end{center}
\end{figure}

\noindent
In Case A, the discussion is a bit more subtle, as far as the experimental constraints are concerned. For mediator masses lower than a hundred keV, the mass scale $\Lambda$ must reach very high values ($\gtrsim 10^{16}~\mrm{GeV}$) to escape experimental exclusion bounds. Still such region of the parameter space is not acceptable since it would lead to a very heavy parameter $\tilde m$. Then for higher masses of the mediator ($ m_{\phi}\gtrsim 300 ~\mrm{keV}$) more reasonable values of $\Lambda$ are allowed, and we are left with lower bounds coming from LEP (mentioned above) and upper bounds on $\Lambda$ arising from CMB dilution and BBN perturbations. Different choices of $\Lambda$ will then lead to different pairs of $(m_{\phi},\tilde m)$, as depicted in Fig.(\ref{Fig:CaseA})

In order to fix ideas, and anticipating results of section \ref{examples}, we indicated in red in the figure the region where $\tilde m\gtrsim m_{\phi}$. This shows clearly, that imposing $m_{\phi} \gtrsim 300~\mrm{keV}$ sets an upper limit for $\Lambda$, giving approximately
\be
\Lambda \lesssim 1000~\mrm{TeV}\,.
\ee
Furthermore, the lower limit $\Lambda \gtrsim 5 ~\mrm{TeV}$ mentioned in section \ref{model} -- still acceptable if there is some strongly coupled hidden sector generating the effective mass scale $\Lambda$ -- imposes an upper limit on the mediator mass, $m_{\phi}\lesssim 50~\mrm{MeV}$\footnote{As mentioned in previous sections, assuming that the effective coupling between the mediator $\phi$ and the photons comes from some perutrbative heavy physics sets a stronger limit on $\Lambda$ leading to masses of the mediator $m_{\phi}\lesssim 5~\mrm{MeV}$.}. One would thus expect from this model that the mediator mass lies in the region
\be
300 ~\mrm{keV}~\lesssim~m_{\phi}~\lesssim~50~\mrm{MeV}\,.
\ee
\section{Relic abundance}

\subsection{State of the art}

\noindent
Computing the relic abundance in models with a very weak annihilation cross section and a keV dark matter  particle
is highly non-standard. Indeed, it is well known from the standard lore that a hot dark matter candidate 
 leads to a relic density
\be
\Omega h^2 \simeq 9.6 \times 10^{-2} \frac{g_{eff}}{g_s(x_f)} \left( \frac{m_s}{1 ~\mrm{eV}} \right)
\ , \label{Eq:hotdm}
\ee

\noindent 
where $g_{eff}$ is the effective number of degrees of freedom of the dark matter candidate and $g_s$ the effective
number of degrees of freedom for the entropy.
Eq.(\ref{Eq:hotdm}) gives $m_s \simeq 5$ eV if one wants to respect PLANCK \cite{PLANCK} limit
$\Omega_{DM} h^2 = 0.1199 \pm 0.0027$.
However, this condition is valid only under the hypothesis that the dark matter is in thermal equilibrium 
with a common temperature $T$ with the thermal bath. In the case of the line signal observed in the clusters,
the cross section necessary to fit the result is far below the classical thermic one 
$\langle \sigma v \rangle_{therm} = 3 \times 10^{-26} \mrm{cm^3 s^{-1}}$. This idea had led previous studies to rule out scalar dark matter candidates lighter than $\mathcal O\left(\mrm{MeV}\right)$ \cite{boehm2}. In fact, the dark bath, composed of
the light mediator $\phi$ and the dark matter $S$ cannot be in equilibrium with the standard plasma. 

\noindent
There are several ways to address this issue. A first possible attempt to solve the problem, proposed in \cite{freezein1} and \cite{freezein2}, is to suppose that the dark matter is
produced through the {\em freeze in} mechanism: the interacting photons annihilate to produce the dark matter in the inverse process
of Fig.(\ref{Fig:feynmanmicro}). Yet it is not possible to get the right relic density in this way since, solving the Boltzmann 
equation in this case would produce too much dark matter. Indeed equilibrium dark matter density would reach quickly a value that would overclose the Universe. 

\noindent
Another way to solve the problem was proposed in \cite{Das:2010ts, Feng:2008mu} where the authors noticed that the condition (\ref{Eq:hotdm})
is not valid anymore if the temperature of the hidden sector $T_h$ is different from the one of the thermal bath $T$.
In this case, one can compute the temperature $T_h$ needed to obtain a 3.56 keV particle respecting the relic abundance 
constraint. Yet, as we will see in what follows, we still need the hidden sector content to be richer in order to provide new dark matter annihilation channels leading to the right relic abundance. This will be done adding to the model a right-handed sterile neutrino.

\subsection{Dark matter annihilation into sterile neutrinos}
\noindent
 One way of solving the lack of annihilation of dark matter described above is to assume that a right-handed sterile neutrino couples directly 
to the mediator scalar particle previously introduced. This would provide another channel for annihilating dark matter which would boost the relic density to its experimental value.

\begin{figure}[htbp]
 \begin{center}
 \includegraphics[width=0.5\linewidth]{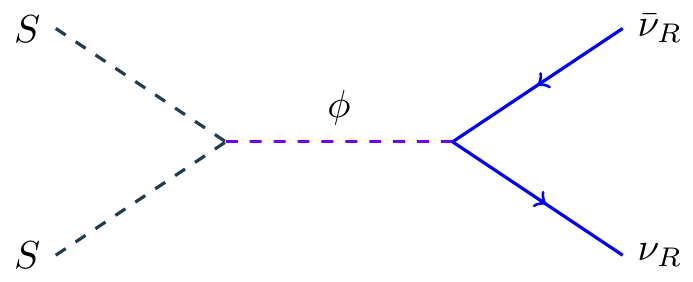}
 \caption{\label{neutrinos} Microscopic model of dark matter  decaying into right-handed sterile neutrinos.}
\end{center}
\end{figure}

\noindent
In a similar fashion than in section \ref{model}, one can add to the usual neutrino interaction terms a Yukawa coupling
 between the field $\phi$ and the sterile neutrino $\nu_R$\footnote{
We use two-component spinor notation in this equation.}
\be
-\mathcal L_{\nu} = \frac{M}{2}\nu_R \nu_R + m_D\nu_L \nu_R + \lambda_{\nu} \phi \nu_R \nu_R+h.c.\,,
\ee

\noindent
After diagonalization of the mass matrix, the sterile neutrino gets mass $m_{st}\simeq M$ while the active one obtains a mass of $m_{act}\simeq m_D^2/M$ via the seesaw mechanism.
\noindent
Such interactions give the following cross section for dark matter annihilation into a pair of sterile neutrinos
\bea
&&\langle \sigma v \rangle_{\nu\nu}  = \frac{\lambda_{\nu}^2\tilde m^2}{8\pi^2}\frac{(m_s^2-M^2)}{m_s^2(4m_s^2-m_{\phi}^2)^2}\sqrt{1-\frac{M^2}{m_s^2}}\,.
\eea

\noindent
Assuming that $M$ is negligible compared to $m_s$  leads to a cross section depending only on $m_s$, $m_{\phi}$ and  $\tilde m$, $\lambda_{\nu}$ :
\bea\label{eq:sigmav}
\langle \sigma v \rangle_{\nu\nu}  &\sim& \frac{\lambda_{\nu}^2}{8\pi^2}\left(\frac{\tilde m}{m_s}\right)^2\frac{m_s^2}{(4m_s^2-m_{\phi}^2)^2}\, \ .
\eea

\subsection{Cosmological constraints on a hidden sector}
\noindent
In such a framework, dark matter and sterile neutrinos can be seen as living together in  a {\em hidden} thermal bath decoupled from the visible sector. Its temperature can be denoted by
\be
T_h \ \equiv \ \xi(t) T\, \ ,
\ee
where in what follows $\xi$  is assumed to be a constant parameter\footnote{See ref. \cite{Feng:2008mu} for a discussion on the validity of this approximation.}.
As described in \cite{Das:2010ts, Feng:2008mu}, such a point of view can have important consequences on the thermal dynamics
 of the hidden sector, since  $\xi$ enters into the Boltzmann equation. Indeed dark matter can decouple while still relativistic or semi-relativistic 
 and  lead to different relic densities, as the parameter $\xi$ takes different values (See Fig.(\ref{Fig:hidden})). 
 Such freedom in the temperature of the hidden sector is 
 yet constrained by astrophysical considerations. In particular, a hidden sector dark matter can freeze out while still being relativistic. It is 
 then important to check that its free streaming length is smaller than typical galactic length scales so that it does not destroy the matter power 
 spectrum. Another constraint, introduced by Tremaine and Gunn in \cite{Tremaine:1979we}, comes from bounding the phase-space density 
 of structures like small dwarf galaxies by statistical quantum mechanics considerations. Both constraints are computed and summarized
  in the case of a $\sim3~\
\mrm{keV}$ dark matter in \cite{Das:2010ts} and lead naturally to a relic density value different from the one of the visible thermal 
bath ($\langle\sigma v\rangle_0\sim 10^{-9} ~\mrm{GeV}^{-2}$), depending on how cold the hidden sector is. We thus get
\be\label{eq:sigmavbounds}
0.015 ~ \langle\sigma v\rangle_0 ~\lesssim ~\langle \sigma v\rangle ~ \lesssim ~0.045 ~\langle\sigma v\rangle_0 ,
\ee

\noindent
the upper limit corresponding to the free streaming constraint whereas the lower one to a strict Tremaine-Gunn bound.
One should also remark that the dark matter is relatively warm ($x_f = m_S/T_f \simeq 2-4$ for points respecting WMAP/PLANCK) in this case.
As a warm candidate, it can elude the classical issues of cold dark matter $i.e.$ the few number of galaxy mergers ($\lesssim 10 \%$)
or the core observed profiles compared to the cusp ones predicted by N-body simulations.

\subsection{Results}
\begin{figure}[htbp]
 \begin{center}
 \includegraphics[width=1.05\linewidth]{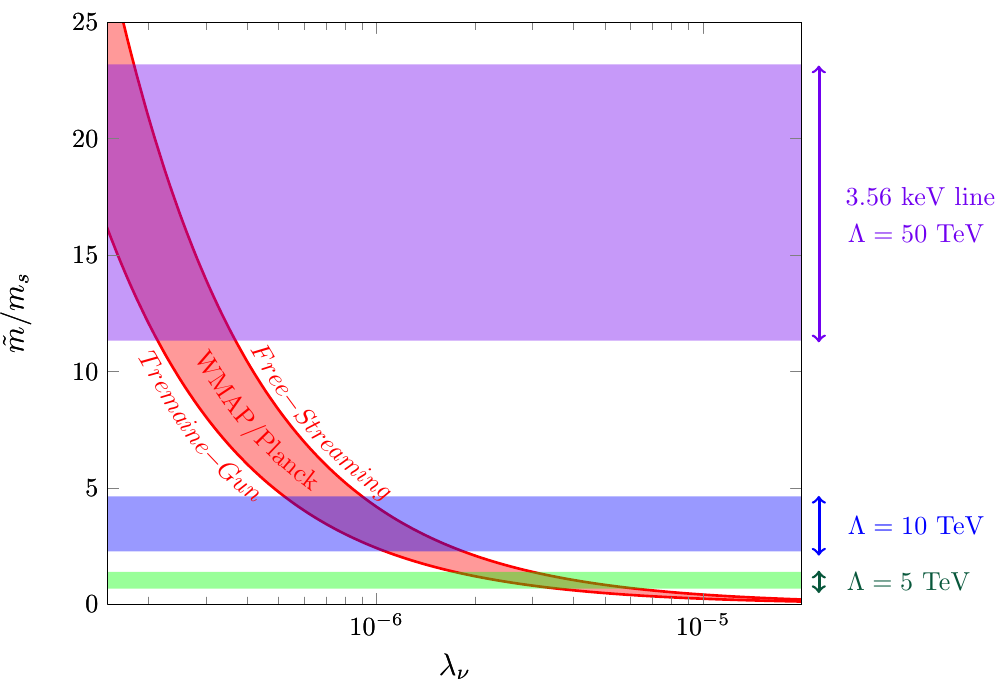}
 \caption{\label{Fig:lambdalambda} {\footnotesize Constraints on the parameter space ($\tilde m$, $\lambda_{\nu}$), for $m_{\phi} = 500 ~\mrm{keV}$; considering cosmological constraints (relic abundance, free streaming and Tremaine-Gunn bounds)
 in the red/dark band and a 3.5 keV excess (blue/green region) for different values of the BSM scale
 $\Lambda$ (see the text for details).}}
\end{center}
\end{figure}

\begin{figure}[htbp]
 \begin{center}
 \includegraphics[width=0.95\linewidth]{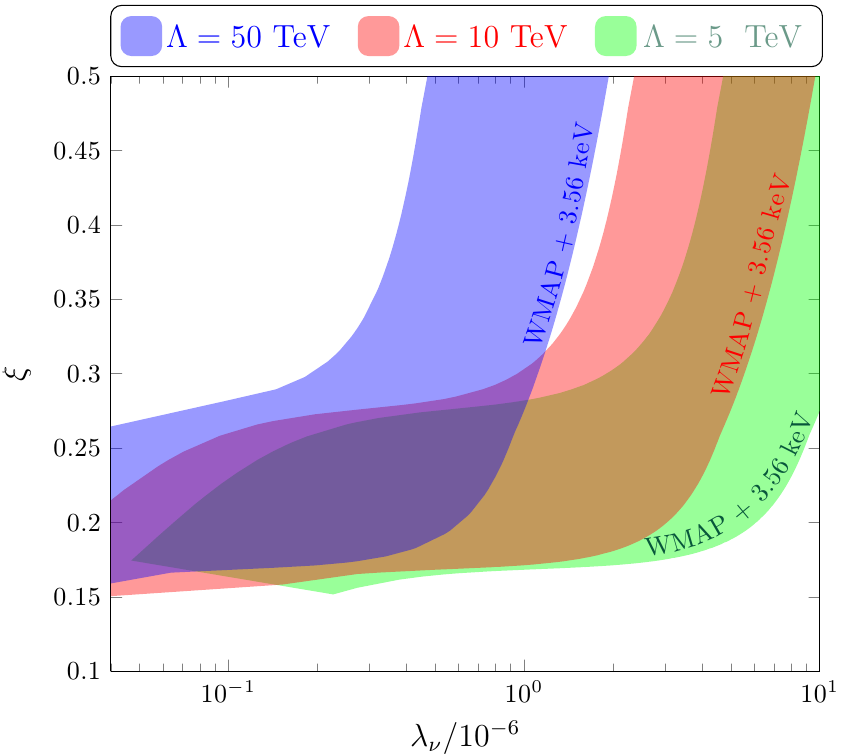}
 \caption{\label{Fig:hidden} {\footnotesize Hidden temperature factor $\xi$ allowed by the flux measurement limits, as a function of the relic density value, for different values of the mass scale $\Lambda$.}}
\end{center}
\end{figure}

\noindent
In the light of section \ref{sec:exp} and equation (\ref{eq:sigmav}) one can now constrain couplings between the mediator $\phi$ to both the dark and the 
visible sector $\tilde m$ and $\lambda_{\nu}$, taking into account that relic density can lie in the region (\ref{eq:sigmavbounds}) exhibited
 in the previous section, as well as imposing constraints on the photons flux measurement. Results are presented in Fig.(\ref{Fig:lambdalambda}) 
 where we show the allowed parameter space imposing cosmological bounds \cite{Das:2010ts} superimposed within the regions fitting the 3.5 keV excess, for 
 $m_\phi=500$ keV. As one can see, a relatively
 large region respects the cosmological bounds and the monochromatic excess. The values of $\lambda_\nu$ are also quite constrained ($\sim 10^{-7}-10^{-5}$, depending on the mass scale $\Lambda$), leading to different values of the hidden sector temperature (as represented in Fig.(\ref{Fig:hidden})). Furthermore it could be interesting 
 in a future work to combine our analysis with neutrino bounds.

\section{An explicit UV model}\label{examples}
\noindent
We give here an explicit example realizing the case A of Section \ref{micro}. The model contains
a scalar $\sigma$ and a pseudoscalar $S$, together with a (set of) fermion(s) $\psi$, with the lagrangian 
\bea
&& {\cal L} = {\cal L}_{kin} + \frac{\mu^2}{2} (\sigma^2 + S^2)  -\frac{\lambda}{4} 
(\sigma^2 + S^2)^2  \nonumber \\
&& - {\bar \psi} (h_1 \sigma + i h_2 S \gamma_5) \psi  +  \frac{\sigma}{\Lambda} F_{\mu \nu} F^{\mu \nu} \ . \label{uv1}
\eea  
The set of fermions $\psi$ are part of the hidden sector,  in particular they are singlets with respect to the Standard Model. 
By defining the complex field $\Phi = \frac{1}{\sqrt 2} (\sigma+i S)$, we can split the lagrangian
(\ref{uv1}) into two parts:
\bea
&& {\cal L} = {\cal L}_0 + {\cal L}_1 \ , {\rm where} \nonumber \\
&& {\cal L}_0 = {\cal L}_{kin} + \mu^2 \Phi^{\dagger} \Phi - \lambda (\Phi^{\dagger} \Phi)^2
- (h_1 {\bar \psi}_L \Phi \psi_R + {\rm h.c.}) \ , \nonumber \\
&&{\cal L}_1  = i (h_1-h_2) \ S \ {\bar \psi} \gamma_5 \psi +  \frac{\sigma}{\Lambda} F_{\mu \nu} F^{\mu \nu}\ . \label{uv2}
\eea
Notice that the whole lagrangian is invariant under the vectorial transformation
\begin{equation}
U(1)_V \quad : \quad \psi \to e^{i \alpha} \psi \quad , \quad \Phi \to \Phi \ ,  \label{uv3} 
\end{equation}
whereas ${\cal L}_0$ is also invariant under the axial transformation
\begin{equation}
U(1)_A \quad : \quad \psi \to e^{i \beta \gamma_5} \psi \quad , \quad \Phi \to  e^{-2 i \beta } \Phi \ .  \label{uv4} 
\end{equation}
The axial transformation is broken by ${\cal L}_1$ and is therefore an exact symmetry of the hidden sector lagrangian in the limit $h_1=h_2$. The symmetry is broken additionally by the coupling to the photons.  
At tree-level, $S$ is massless since it is the (pseudo)Goldstone boson of the axial $U(1)_A$ symmetry. 
For $\mu^2 > 0$ there is a symmetry breaking vacuum, with $\langle \sigma \rangle = v$ and
$v^2 = \frac{\mu^2}{\lambda}$. By expanding around the minimum one also finds at tree-level
\be
m_{\phi}^2 = 2 \mu^2 \quad , \quad {\tilde m} = \sqrt{\frac{\lambda}{2}} \ m_{\phi} \ . \label{uv5} 
\ee
At the one-loop level, there is a quantum correction to the potential $V_1 = V_1 (h_1^2 \sigma^2 + h_2^2
S^2)$, which generates a mass for the pseudo-goldstone boson $S$, proportional to
$m_S^2 \sim O(\frac{h_2^2-h_1^2}{16 \pi^2}) m_{\phi}^2$, which has a one-loop suppression with respect to the mediator mass $\phi$. This model is an example of a UV embedding of case A of Section \ref{micro}, in which one naturally expects ${\tilde m} \leq m_{\phi}$ and selects therefore natural regions in the parameter space in Section \ref{micro}. 
\section*{Conclusions}
\noindent
In this work, we have shown that a keV scalar dark matter can be the main constituent of the matter of the universe, producing monochromatic X-ray signals that can be fitted with the recently claimed events of a 3.5 keV line in nearby clusters of galaxies. Moreover, we know that for a warm dark matter mass of order of a keV, free streaming produces a cutoff in the linear fluctuation power spectrum at a scale corresponding to dwarf galaxies and can fit observations for $m_s \gtrsim 1.5$ keV \cite{Lovell:2013ola}. 
 We have shown that astrophysical, collider and relic density constraints are more difficult
to accomodate. They are however possible to satisfy for certain values of the mediator mass 
$m_{\phi}$ and scale $\Lambda $ of the  couplings between the mediator and the photon
$300 ~\mrm{keV}~\lesssim~ m_{\phi}~\lesssim ~50 ~\mrm{MeV}$ and $5 ~\mrm{TeV}~\lesssim ~\Lambda~ \lesssim ~1000~\mrm{TeV}$. These values can conversely be considered as a prediction of our setup of keV dark matter models leading to X-ray monochromatic lines. 
We also showed that as pointed out recently by \cite{Papucci:2014iwa}, the study of $pure$ effective models -- as it is often done in the literature -- misses important quantitative effects. Indeed, we could reach a cross section in agreement with the analysis of XMM Newton data only through the building of a microscopic model and the exchange of a light scalar. Finally, it should be interesting to look at the observability of such a light meson. This task is far beyond the scope of our paper but is under investigation.


\noindent {\bf Acknowledgements. }  The authors would like  to thank M. Goodsell and  A. Ringwald for very useful discussions.  E.D. and L.H. would like to thank the Alexander von Humboldt foundation and DESY-Hamburg for support and hospitality during
the final stage of the project and  Y.M. would like to thank K. Tarachenko for constant support.
This  work was supported by
the French ANR TAPDMS {\bf ANR-09-JCJC-0146}, the Spanish MICINN's
Consolider-Ingenio 2010 Programme  under grant  Multi-Dark {\bf CSD2009-00064},  the European Union FP7 ITN INVISIBLES (Marie
Curie Actions, PITN- GA-2011- 289442) and  the ERC advanced grants  
 MassTeV and  Higgs@LHC. 
\section*{Appendix}

\noindent
For completeness, we present in this appendix the main alternatives to the Lagrangian (\ref{Eq:lagrangian}). They do not affect however our conclusions.

\subsection*{Fermionic dark matter}
\subsubsection*{Effective operator approach}

 With the same philosophy used in section \ref{Sec:effective}, we could consider a fermionic dark matter $\chi$ interacting with photons through effective operators of the form
\be
\frac{\bar \chi \chi}{\Lambda^3} F_{\mu \nu} F^{\mu \nu}\,,\,\,\frac{\bar \chi \gamma_5 \chi}{\Lambda^3} F_{\mu \nu} \tilde F^{\mu \nu} \ , 
\ee
\begin{figure}
\begin{center}
 \includegraphics[width=0.40\linewidth]{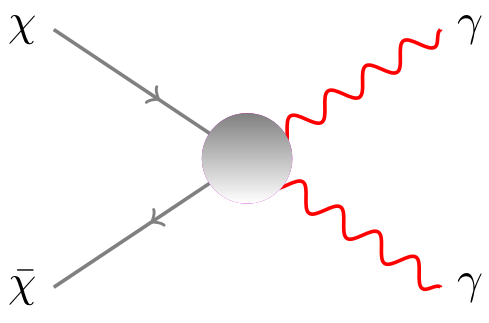}
 \caption{{\footnotesize Effective diagram for dark matter annihilation in the fermionic case}}
\label{Fig:feynmanmicrofermion}
\end{center}
\end{figure}
\noindent
where $\tilde F^{\mu\nu} =  \frac{1}{2}\epsilon^{\mu\nu\rho\sigma}F_{\rho\sigma}$. The obtained cross sections are, respectively
\be
\langle \sigma v \rangle_{\gamma \gamma}^{eff} \,=\, \frac{8 m_{\chi}^4v^2}{\pi \Lambda^6}\,\,,\,\,\,\,\,\frac{8 m_{\chi}^4}{\pi \Lambda^6}\, .
\label{Eq:sigmaeffective2}
\ee

\noindent
The constraints (\ref{Eq:constraints}) give then
\be
0.061 < \Lambda < 0.078 \,~~,~~~\,\,0.091 < \Lambda < 0.12 ~~\,\,(\mrm{GeV})
\label{Eq:resultseffective2}
\ee

\noindent
 It is clear then  that in the fermionic case these values are even more incompatible with experimental constraints coming from colliders.
\subsubsection*{Microscopic Model}
{\bf Scalar Mediator.}
\noindent
Following the discussion of section \ref{micro} but working in the alternative frame of a fermionic dark matter (see Fig.(\ref{Fig:feynmanmicrofermion}))
particle interacting with the standard model via a scalar portal, one can write the following lagrangian

\bea
{\cal L}_{eff} = g_{\chi} \phi \bar \chi \chi + \frac{\phi}{\Lambda}F_{\mu\nu}F^{\mu\nu}\, \ .
\nonumber
\eea
\noindent
One then gets the annihilation cross section

\be
\langle \sigma v \rangle_{\gamma \gamma}^{micro} = \frac{2g_{\chi}^2 m_{\chi}^4v^2}{ \pi \Lambda^2 (4 m_{\chi}^2 - m_\phi^2)^2} \ . 
\ee
which in what follows is fixed in order to fit the X-ray monochromatic signal.
\begin{figure}
\begin{center}
 \includegraphics[width=0.65\linewidth]{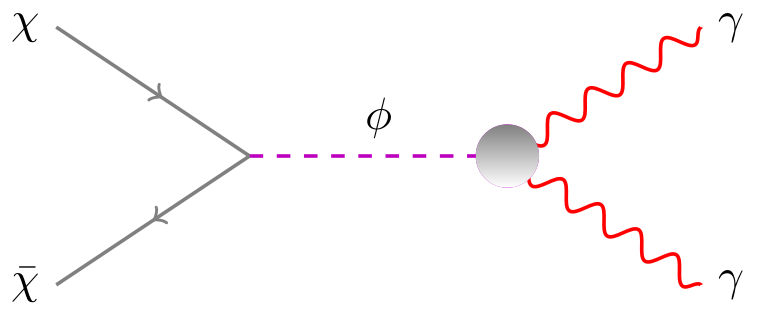}
 \caption{{\footnotesize Microscopic diagram for dark matter annihilation in the fermionic case}}
\label{Fig:feynmanmicrofermion}
\end{center}
\end{figure}

\noindent
Furthermore, one can still differentiate between the two cases where the mass of the mediator is heavier or lighter than the dark matter one. 

\bea
\text{{\bf Case A}} &:&  ~m_\phi \gtrsim m_{\chi}~\text{{(Heavy Mediator)}}  
\nonumber
\\
\nonumber\\
m_\phi \simeq (0.97 &-& 1.4)\left( \frac{2 g_{\chi}^2v^2}{\pi} \right)^{1/4} \left( \frac{m_{\chi}}{3.5 ~\mrm{keV}} \right) \sqrt{\frac{1 ~\mrm{TeV}}{\Lambda}}
~\mrm{MeV}
\nonumber
\\
\label{Eq:heavy2}\\
\text{{\bf Case B}} &:& ~ m_\phi \lesssim m_{\chi}~\text{{(Light Mediator)}} 
\nonumber
\\
\nonumber\\
&&\frac{g_{\chi}^2v^2}{\Lambda^2} \sim  4.2 \times 10^{-14} ~\mrm{GeV^{-2}}
\label{Eq:line2}
\eea

\noindent
The cross section is  velocity suppressed here and leads, for a given value of the mediator mass and of the coupling 
 $g_{\chi}$, to smaller values of the mass scale $\Lambda$. 
 For example, for $g_{\chi}=1$ in case B, one gets  $\Lambda \sim  1.4\times 10^6 \,\mrm{GeV}$. 

\vspace{4pt}
{\bf Pseudo-Scalar Mediator.}
\noindent
Another option is to make the dark sector communicate with the standard model through the exchange of a pseudo-scalar particle. The lagrangian is given by
\bea
{\cal L}_{eff}= g_{\chi} \phi \bar \chi \gamma_5 \chi + \frac{\phi}{\Lambda}F_{\mu\nu}\tilde F^{\mu\nu}\,,
\nonumber
\eea

and the annihilation cross section is

\be
\langle \sigma v \rangle_{\gamma \gamma}^{micro} = \frac{2g_{\chi}^2 m_{\chi}^4}{ \pi \Lambda^2 (4 m_{\chi}^2 - m_\phi^2)^2} \ , 
\ee

which is now not velocity suppressed anymore.
\bea
\text{{\bf Case A}} &:&  ~m_\phi \gtrsim m_{\chi}~\text{{(Heavy Mediator)}} 
\nonumber
\\
\nonumber\\
m_\phi \simeq (0.97 &-& 1.4)\left( \frac{2 g_{\chi}^2}{\pi} \right)^{1/4} \left( \frac{m_{\chi}}{3.5 ~\mrm{keV}} \right) \sqrt{\frac{1 ~\mrm{TeV}}{\Lambda}}
~\mrm{MeV}
\nonumber
\\
\label{Eq:heavy3}\\
\text{{\bf Case B}} &:& ~ m_\phi \lesssim m_{\chi}~\text{{(Light Mediator)}}
\nonumber
\\
\nonumber\\
&&
\frac{g_{\chi}^2}{\Lambda^2} \sim  4.2 \times 10^{-14} ~\mrm{GeV^{-2}}
\label{Eq:line3}
\eea
\noindent In case B fitting the X-ray signal requires therefore $\Lambda \sim 5\times10^6  \,\mrm{GeV}$.

\noindent
In both the cases of scalar and pseudo-scalar mediators, as is the case in the rest of the paper, case B is always excluded, regarding to the HB experimental bounds. As far as the Case A is concerned, the discussion is similar and $g_{\chi}$ plays the same role as $\tilde m$. However the needed values of $\Lambda$ are different compared to the scalar dark matter : $\Lambda \lesssim 10 ~\mrm{TeV}$ for Eq.(\ref{Eq:heavy2}) and $\Lambda \lesssim 100 ~\mrm{TeV}$ for Eq.(\ref{Eq:heavy3}).

\end{document}